\documentclass[%
 reprint,
groupedaddress,
amsmath,amssymb,
aps,
pra,
]{revtex4-1}
\usepackage{graphicx}
\usepackage{dcolumn}
\usepackage{bm}
\usepackage{float}
\usepackage{multirow}

\usepackage[dvipdfm,colorlinks,linkcolor=blue, urlcolor=blue, anchorcolor=blue, citecolor=blue]{hyperref}
\begin{document}


\title{Resonator-qutrits quantum battery}
\author{Fang-Mei Yang}%
\affiliation{College of Physics and Electronic Engineering, Northwest Normal University, Lanzhou, 730070, China}
\author{Fu-Quan Dou}
\email{doufq@nwnu.edu.cn}
\affiliation{College of Physics and Electronic Engineering, Northwest Normal University, Lanzhou, 730070, China}

\begin{abstract}
Quantum batteries (QBs) are energy storage and transfer microdevices that open up new possibilities in energy technology. Here, we derive a resonator-qutrits quantum battery (QB) model consisting of a multi-modes resonator and $N$ superconducting transmon qutrits. We investigate the charging and self-discharging performance of the QB and discuss the roles of quantum coherence and quantum entanglement. The results show that environment noise is not always detrimental for QB systems. The QB with efficient charging, stable energy-storage and slow self-discharging processes can be realized by considering the dephasing noise and manipulating the energy gap. We find that the charing energy is positively related to coherence and entanglement while the stable energy and the self-discharing energy are negatively related to coherence. The phenomenon of the vanishing entanglement corresponds to the dynamic decoupling behavior of the QB's steady states. Our results provide a way to realize many-body QBs on superconducting circuits platform.
\end{abstract}
\maketitle
\section{Introduction}
Motivated by the ongoing miniaturization of electronic equipments and the rapid development of quantum thermodynamics in micro- and mesoscopic quantum systems, Alicki and Fannes first introduced the notion of quantum battery (QB) to temporarily store energy and suggested that entangling unitary operations can extract more work than local ones \cite{PhysRevE.87.042123}. Subsequently, it was shown that there exists some protocols for optimal work extraction actually without creating any entanglement, at the cost of generically requiring more operations \cite{PhysRevLett.111.240401,Giorgi2015}. Further researches uncovered that entanglement may have a destructive effect on the performance of quantum batteries (QBs) \cite{PhysRevB.104.245418,PhysRevLett.122.047702,PhysRevLett.129.130602}, whereas coherence appears to be a potential resource \cite{PhysRevE.102.052109,PhysRevE.102.042111,PhysRevLett.129.130602,Gumberidze2019,PhysRevLett.125.040601,PhysRevA.99.052105,PhysRevA.105.062203,PhysRevLett.127.100601}. In recent years, how to exploit quantum resources like entanglement and coherence to outperform its classical counterpart has been a major focus in the quantum thermodynamics realm, but the couplings among different quantum resources lead to the problem intractable and complicated \cite{PhysRevA.104.L030402,PhysRevApplied.14.024092,PhysRevLett.125.236402,PhysRevLett.128.140501,Goold2016,Millen2016,PhysRevLett.118.150601,PhysRevLett.129.130602,Guryanova2016}.

Nowadays, the development of QBs is still in their infancy and there are many challenges that need to be adequately solved before such quantum devices can be implemented in practice. Besides the widely concerned efficient charging \cite{Crescente2020,PhysRevA.97.022106,PhysRevA.106.022618,Shaghaghi2022,PhysRevA.107.042419,PhysRevA.107.032218,PhysRevA.102.060201,PhysRevB.98.205423,PhysRevB.102.245407,PhysRevLett.120.117702,PhysRevLett.125.236402,PhysRevLett.122.210601,PhysRevB.105.115405,PhysRevA.106.032212,PhysRevResearch.4.013172,dou2022charging,dou2023analytically,dou2023three} and stable energy-storage processes \cite{PhysRevApplied.14.024092,PhysRevE.100.032107,PhysRevE.101.062114,PhysRevResearch.2.013095,Moraes2021,Rosa2020,PhysRevE.104.044116,PhysRevA.103.033715,PhysRevA.107.023725,Dou2020,Dou2021}, it is also important to highlight the self-discharging process characterized by the decay of energy when no consumption hubs are connected to QBs. How to against the undesired effects and boost the ability of QBs in storing energy during the self-discharging process is a key task \cite{PhysRevE.100.032107,Kamin2020,PhysRevE.103.042118,PhysRevE.105.054115}. At present, 
some protocols have been purposed to preserve the energy stored in QBs for long times, e.g., considering spectrum engineering \cite{PhysRevE.100.032107} and non-Markovian effects \cite{Kamin2020}, introducing local field fluctuation described by disorder term in the system Hamiltonian \cite{PhysRevE.103.042118}, manipulating coherence of the initial states and geometry of the intracell connections \cite{PhysRevE.105.054115}.

Experimentally, several progresses have exhibited some advantages of QBs. 
The first realization of QB, using an organic semiconductor as an ensemble of two-level systems coupled to a microcavity, shows that the power density of Dicke QB is up to 67 kW/kg which far exceeds the power densities of existing lithium-ion batteries \cite{metzler2023quantum,quach2020}.
Up to now, QBs can be promisingly implemented on many physical platforms \cite{PhysRevLett.124.130601,Cruz2022,gemme2022ibm,quach2020,PhysRevA.106.042601,wenniger2022coherence,PhysRevA.107.L030201,PhysRevLett.131.240401,Hu2022,Zheng2022,Ge2023}, ranging from organic semiconductor \cite{quach2020} and semiconductor quantum dot \cite{wenniger2022coherence} coupled to optical microcavities, nuclear spin systems with a star-topology configuration \cite{PhysRevA.106.042601}, photonic systems using polarizations of single photons \cite{PhysRevA.107.L030201,PhysRevLett.131.240401}, to superconducting circuits systems operating at milli-Kelvin temperatures \cite{Hu2022,Zheng2022,Ge2023}. Compared to other platforms, superconducting circuits are widely concerned since its flexibility, adjustability, scalability and strong coupling with external fields \cite{You2005,H.2013,RevModPhys.85.623,Gu2017,PRXQuantum.2.040202}. In 2019, Santos et al. have proposed that superconducting circuits is an effective platform to realize QB \cite{PhysRevE.100.032107} and the viewpoint has been confirmed in subsequent experiments which generally consist of one superconducting qutrit driven by classical fields \cite{Hu2022,Zheng2022,Ge2023}. Remarkably, the semi-classical QB model designed on superconducting circuits can ensure a stable charged state by utilizing the stimulated Raman adiabatic passage, but has limited charging energy and fast self-discharging speed. Therefore, the exploration of many-body QB with efficient charging, stable energy-storage and slow self-discharging processes on superconducting circuits platform is necessary and meaningful.

In this paper, we focus on a superconducting circuit system composed by a one-dimensional transmission line resonator and $N$ coupled transmons. Under the condition that the anharmonicity of transmon is smaller than the transmon-resonator detuning, we derive the general Hamiltonian of this superconducting circuit system and define three QBs, including the one-mode or two-modes resonator-qutrit QBs and the resonator-qutrits QB. We investigate the performance of these QBs in the presence of the resonator decay and the qutrit decoherence, then discuss the roles of quantum entanglement and quantum coherence during their charging and self-discharging processes.

The paper is organized as follows. In Sec. \ref{section2} we derive the Hamiltonian of a superconducting circuit system and define a general QB model. In Sec. \ref{section3} we investigate the performance of the QB in the presence of the resonator decay and the qutrit decoherence. The roles of quantum entanglement and quantum coherence during their charging and self-discharging processes are discussed in Sec. \ref{section4}. Finally, a brief conclusion is given in Sec. \ref{section5}.
\section{Model and Hamiltonian} \label{section2}
\subsection{The superconducting circuit system}
Our previous work has proposed a superconducting circuit system and we have derived the Hamiltonian of the resonator-qubits system only considering the lowest mode of the resonator in the small resonator-transmon detuning regime \cite{PhysRevA.107.023725}. However, when the transmon anharmonicity is small compared to the resonator-transmon detuning, the transitions between states $|m+1\rangle$ and $|m\rangle$ are almost equally likely to be excited by the resonator, causing non-negligible leakage to higher excited states \cite{PhysRevA.86.013814,PhysRevA.88.052330,PhysRevLett.114.010501,Pirkkalainen2013}. In such a condition, we expect to derive a more general light-matter interaction model based on superconducting circuits system.

The classical Hamiltonian of the superconducting circuit system, which consists of a one-dimensional transmission line resonator and $N$ capacitively coupled transmons, is described by \cite{PhysRevA.107.023725}
\begin{eqnarray}
H^{cl}=H^{cl}_{r}-\sum_{i=1}^{N}C_{c}\dot{\Phi}_{r}\dot{\Phi}_{i}+\frac{1}{2}\vec{\dot{\Phi}}\mathcal{C}\vec{\dot{\Phi}}^{T}-\sum_{i=1}^{N}2E_{J}cos\delta,
\end{eqnarray}
where $H^{cl}_{r}$ is the classical Hamiltonian of the transmission line resonator, $\vec{\dot{\Phi}}=(\dot{\Phi}_{1},\dot{\Phi}_{2},...,\dot{\Phi}_{N})$ is the vector of voltage, $\dot{\Phi}_{i} (i=1,\cdots N)$ and $\dot{\Phi}_{r}$ represent the voltage of the $i$th transmon and the resonator. $E_{J}$ is Josephson energy of each Josephson junction, $\delta$ is the phase difference between Josephson junctions consisting transmon and $\mathcal{C}$ is the capacitance matrix defined as
\begin{equation*}
\mathcal{C}=
	\begin{bmatrix}
    C_{0}+C&-C& & & &\\
	-C&C_{0}+2C&-C& &\\
	 &-C&C_{0}+2C&-C& &\\
	 & &-C&\ddots&\ddots&\\
     & & &\ddots& &
	\end{bmatrix}.
\end{equation*}
Here, $C_{0}=2C_{J}+C_{B}+C_{g}+C_{c}$ and $C_{J}, C_{B}, C_{g}, C_{c}, C_{r}, C$ are capacitance in the superconducting circuit.

According to the quantization procedure of the transmon \cite{PhysRevLett.129.087001,PhysRevA.94.033850,PhysRevLett.127.237702} and the transmission line resonator \cite{PhysRevA.69.062320,RevModPhys.93.025005}, we express
\begin{eqnarray}
\begin{split}
&\vec{\dot{\Phi}}^{T}=2e\mathcal{C}^{-1}(\vec{n}-n_{g})^{T},\\
&\dot{\Phi}_{r}=\sum_{k}\sqrt{\hbar\omega_{rk}/(C_{r}+NC_{c})}(a_{k}+a_{k}^{\dag}),
\end{split}
\end{eqnarray}
where $\vec{n}=(n_{1},n_{2},...,n_{N})$ is the vector of the Cooper pair number, $n_{i}$ is the number of Cooper pairs transferred between Josephson junctions and $n_{g}$ is the number of gate charge. The quantized Hamiltonian of the resonator $H_{r}=\sum_{k}\hbar\omega_{rk}a_{k}^{\dag}a_{k}$, with the frequency $\omega_{rk}$ and the creation (annihilation) operator $a_{k}^{\dag}~(a_{k})$ of the $k$th harmonic oscillator constituting the resonator.  Thus, the quantized Hamiltonian of the whole system can be written as
\begin{eqnarray}
\begin{split}
\label{quHam}
H=&\sum_{k}\hbar\omega_{rk}a_{k}^{\dag}a_{k}+\sum_{i=1}^{N}\left[2e^{2}\mathcal{C}_{ii}^{-1}(n_{i}-n_{g})^{2}-2E_{J}cos\delta\right]\\
&+\sum_{i<j}^{N}4e^{2}\mathcal{C}_{ij}^{-1}(n_{i}-n_{g})(n_{j}-n_{g})\\
&-\frac{2eC_{c}}{C_{0}}\sum_{k}\sqrt{\frac{\hbar\omega_{rk}}{C_{r}+NC_{c}}}(a_{k}+a_{k}^{\dag})\sum_{i=1}^{N}(n_{i}-n_{g}).
\end{split}
\end{eqnarray}
Defining $E_{C}=e^{2}/2C_{0}, \beta=C/(C_{0}+C)$ and substituting $\mathcal{C}_{ii}^{-1}\approx1/C_{0}, \mathcal{C}_{ij}^{-1}\approx\beta^{|i-j|}/C_{0}$ into Eq. (\ref{quHam}), the above Hamiltonian becomes
\begin{eqnarray}
\begin{split}
H=&\sum_{k}\hbar\omega_{rk}a_{k}^{\dag}a_{k}+\sum_{i=1}^{N}\left[4E_{C}(n_{i}-n_{g})^{2}-2E_{J}\cos\delta\right]\\
&+\sum_{i<j}^{N}8E_{C}\beta^{|i-j|}(n_{i}-n_{g})(n_{j}-n_{g})\\
&-\frac{2eC_{c}}{C_{0}}\sum_{k}\sqrt{\frac{\hbar\omega_{rk}}{C_{r}+NC_{c}}}(a_{k}+a_{k}^{\dag})\sum_{i=1}^{N}(n_{i}-n_{g}).
\end{split}
\end{eqnarray}
It is noted that the Josephson energy $E_{J}$ acts as a strong ``gravitational force" on the rotor, effectively restricting the angle $\delta$ to small values around zero \cite{PhysRevA.76.042319}. This motivates the expansion of the cosine terms up to fourth order, then the Hamiltonian can be expressed as
\begin{eqnarray}
\begin{split}
\label{quHam1}
H=&\sum_{k}\hbar\omega_{rk}a_{k}^{\dag}a_{k}+\sum_{i=1}^{N}\left[\omega_{q}b_{i}^{\dagger}b_{i}-\frac{E_{C}}{12}\left( b_{i}^{\dagger}+b_{i}\right)^{4}\right]\\ &-\frac{\omega_{q}}{2}\sum_{i<j}\beta^{|i-j|}(b_{i}-b_{i}^{\dagger})(b_{j}-b_{j}^{\dagger})\\
&+I\sum_{k}\sqrt{\frac{\omega_{q}\omega_{rk}E_{C}C_{c}^2}{e^{2}(C_{r}+NC_{c})}}(a_{k}+a_{k}^{\dag})\sum_{i=1}^{N}(b_{i}-b_{i}^{\dagger}),\\
\end{split}
\end{eqnarray}
where $b_{i}^{\dagger}~(b_{i})$ is the creation (annihilation) operator of the $i$th transmon, $I$ represents the imaginary unit, $\delta=2\sqrt{E_{C}/\omega_{q}}(b_{i}+b_{i}^{\dagger}), n_{i}-n_{g}=-I\sqrt{\omega_{q}/E_{C}}(b_{i}-b_{i}^{\dagger})/4$, and $\omega_{q}=\sqrt{16E_{C}E_{J}}$ is the frequency of the transmon.

In the following, we only consider three lowest energy levels of the transmon since the main leakage out of the qubit basis comes from the third energy level \cite{PhysRevA.94.063861,PhysRevA.98.052314,PhysRevA.107.023725}. The Hamiltonian (\ref{quHam1}) can be truncated to the three levels (including the ground state $|0\rangle$, the first excited state $|1\rangle$ and the second excited state $|2\rangle$) and written as
\begin{eqnarray}
\begin{split}
H=&\sum_{k}\hbar\omega_{rk}a_{k}^{\dag}a_{k}+\sum_{i=1}^{N}S^{z}_{i}\\
&+\frac{I^{2}\omega_{q}}{2}\sum_{i<j}\beta^{|i-j|}(S^{-}_{i}-S^{+}_{i})(S^{-}_{j}-S^{+}_{j})\\
&+I\sum_{k}\sqrt{\frac{\omega_{q}\omega_{rk}E_{C}C_{c}^2}{e^{2}(C_{r}+NC_{c})}}(a_{k}+a_{k}^{\dag})\sum_{i=1}^{N}(S^{-}_{i}-S^{+}_{i}),
\end{split}
\end{eqnarray}
where the annihilation operator of the transmon $b$ ($b\equiv|0\rangle\langle 1|+\sqrt{2}|1\rangle\langle 2|+\sqrt{3}|2\rangle\langle 3|+...$\cite{PhysRevA.91.043846}) can be treated as $S^{-}_{i}=|0\rangle\langle1|+\sqrt{2}|1\rangle\langle2|$. $S^{+}_{i}=|1\rangle\langle0|+\sqrt{2}|2\rangle\langle1|, S^{z}_{i}=\omega_{0}|0\rangle\langle0|+\omega_{1}|1\rangle\langle1|+\omega_{2}|2\rangle\langle2|$ and $\omega_{m}$ is the frequency of the state $|m\rangle$. We ignore the long-range interaction and the final quantized Hamiltonian of the resonator-qutrits system takes the following form
\begin{eqnarray}
\begin{split}
H=&\sum_{k}\hbar\omega_{rk}a_{k}^{\dag}a_{k}+\sum_{i=1}^{N}S_{i}^{z}\\
&+I^{2}J\sum_{i=1}^{N-1}(S^{-}_{i}-S^{+}_{i})(S^{-}_{i+1}-S^{+}_{i+1})\\
&+I\sum_{k}g_{k}(a_{k}+a_{k}^{\dag})\sum_{i=1}^{N}(S^{-}_{i}-S^{+}_{i}),
\end{split}
\end{eqnarray}
where $\omega_{rk}=k\pi/\sqrt{L_{r}(C_{r}+NC_{c})}$ is the frequency of the resonator, $J=\omega_{q}\beta/2$ is the nearest neighbor interaction strength between the qutrits, and $g_{k}=\sqrt{\omega_{q}\omega_{rk}E_{C}C_{c}^2/(e^{2}(C_{r}+NC_{c}))}$ is the coupling strength between the resonator and the qutrits.
\subsection{The resonator-qutrits QB model}
For the convenience of numerical simulation, we consider that the transmon qutrits are coupled to the two modes ($k=1, 2$) of the transmission line resonator and the Hamiltonian of the resonator-qutrits system can be simplified to (hereafter we set $\hbar=1$)
\begin{eqnarray}
&H=H_{r}+H_{q}+H_{r-q},
\end{eqnarray}
where
\begin{eqnarray}
\begin{split}
\label{RQHam}
&H_{r}=\sum_{k=1}^{2}\omega_{rk}a_{k}^{\dag}a_{k},\\
&H_{q}=\sum_{i=1}^{N}S_{i}^{z}+I^{2}J\sum_{i=1}^{N-1}(S^{-}_{i}-S^{+}_{i})(S^{-}_{i+1}-S^{+}_{i+1}),\\
&H_{r-q}=Ig(a+a^{\dag})\sum_{i=1}^{N}(S^{-}_{i}-S^{+}_{i}).
\end{split}
\end{eqnarray}
Here we set $g=\sqrt{\omega_{q}\omega_{r}E_{C}C_{c}^2/(e^{2}(C_{r}+NC_{c}))}$, $\omega_{r}=2\pi/\sqrt{L_{r}(C_{r}+NC_{c})}$. $H_{r}, H_{q}, H_{r-q}$ are the Hamiltonian of the transmission line resonator, the transmon qutrits with nearest neighbor interactions and the resonator-qutrits coupling, respectively.

We define that the $N$ capacitively coupled transmon qutrits $H_{q}$ plays the role of battery and the resonator $H_{r}$ as a charger. In real scenarios, the QB systems are regarded as open systems due to the inevitable interactions with complex environments. Therefore we mainly explore the charging and self-discharging performance of our QB with three relevant environmental effects.

When the coupling between the resonator and the qutrits is switched on, the charging process immediately starts. The charging dynamics of the QB is described by the Lindblad master equation
\begin{equation}
\label{Ceq}
\dot{\rho_{c}}(t)=I[\rho_{c},H]+\mathbb{L}_{a}[\rho_{c}]+\mathbb{L}_{rel}[\rho_{c}]+\mathbb{L}_{dep}[\rho_{c}],
\end{equation}
where the superoperators $\mathbb{L}_{a}[\bullet]$, $\mathbb{L}_{rel}[\bullet]$ and $\mathbb{L}_{dep}[\bullet]$ describe three relevant environmental effects including the resonator decay, the qutrit relaxation and the qutrit dephasing processes, respectively. $\mathbb{L}_{a}[\bullet]$, $\mathbb{L}_{rel}[\bullet]$ and $\mathbb{L}_{dep}[\bullet]$ can be written as
\begin{eqnarray}
\begin{split}
\label{superoperators}
&\mathbb{L}_{a}[\bullet]=\kappa\left[ a\bullet a^{\dag}-\frac{1}{2}\{a^{\dag}a,\bullet\}\right],\\
&\mathbb{L}_{rel}[\bullet]=\sum_{m=1,2}\Gamma_{nm}\left[S_{nm}\bullet S_{mn}-\frac{1}{2}\{S_{mm},\bullet\}\right],\\
&\mathbb{L}_{dep}[\bullet]=\sum_{m=1,2}\Gamma_{mm}\left[S_{mm}\bullet S_{mm}-\frac{1}{2}\{S_{mm},\bullet\}\right].
\end{split}
\end{eqnarray}
In the above superoperators, $S_{mn}=\sum_{i=1}^{N}\sigma_{i}^{mn}$,~$\sigma_{i}^{mn}=|m\rangle\langle n|$,~$n=m-1$~($m=1,2$),~$\rho_{c}(t)=|\psi_{c}(t)\rangle\langle \psi_{c}(t)|$ is the density matrix of the QB system during the charging process. These three relevant environmental effects corresponding to the resonator decay rate $\kappa$, the qutrit relaxation rate $\Gamma_{01}, \Gamma_{12}$ and the qutrit dephasing rate $\Gamma_{11}, \Gamma_{22}$, respectively.

We prepared the initial state of battery and charger in the ground state $|G\rangle=|0\rangle^{\bigotimes N}$ and the $n_{r}$ photons' Fock state $|n_{r}\rangle$. Thus, the initial state of the QB system during the charging process is as follows
\begin{equation}
|\psi_{c}(0)\rangle=|G\rangle\otimes|n_{r}\rangle.
\end{equation}
The energy in the battery can be defined in terms of the mean of Hamiltonian $H_{q}$ at time $t$, i.e.,
\begin{equation}
E_{c}(t)=\rm{Tr}[H_{q}\rho_{q}(t)],
\end{equation}
where $\rho_{q}(t)=\rm{Tr}_{r}[\rho_{c}(t)]$ is the reduced density matrix of the battery. The stored energy is the difference of energy between the final and initial battery sates
\begin{equation}
\Delta E_{c}(t)=E_{c}(t)-E_{c}(0),
\end{equation}
where $E_{c}(0)=0$ is the ground-state energy of the battery. The average charging power is given by
\begin{equation}
P_{c}(t)=\Delta E_{c}(t)/t.
\end{equation}
In addition to the stored energy $\Delta E_{c}$ and the average charging power $P_{c}$, equally important is assessing the stable energy $E_s$ and the maximum power $P_{max}$ of the QB, which can be quantified as
\begin{equation}
E_s=\Delta E_{c}(\infty), P_{max}=\max[P_{c}(t)].
\end{equation}
For open systems, the logarithmic negativity provides a convenient measure of entanglement \cite{RevModPhys.80.517,PhysRevApplied.14.024092}. The entanglement between the charger and the battery is defined by the trace norm as
\begin{equation}
S_{c}(t)=\log_{2}\|\rho_{q}(t)\|.
\end{equation}
One of the most common measures of coherence for quantum states is the $l_{1}$ norm of coherence measuring the overall magnitude of off-diagonal elements \cite{PhysRevLett.113.140401}, which we express as
\begin{equation}
C_{c}(t)=\sum_{i\neq j}\mid\rho_{q}(t)\mid.
\end{equation}

When the battery is disconnected from the charger and any consumption hub, the resonator-qutrits QB begins to self-discharge due to the inevitable interaction between the qutirts and the environment. Here we consider the qutrit decoherence process (a combination of relaxation and dephasing process), the self-discharging dynamics of the QB is obtained by solving the following equation
\begin{equation}
\label{disCeq}
\dot{\rho_{d}}(t)=I[\rho_{d},H_{q}]+\mathbb{L}_{rel}[\rho_{d}]+\mathbb{L}_{dep}[\rho_{d}].
\end{equation}
The form of the superoperators $\mathbb{L}_{rel}[\bullet]$ and $\mathbb{L}_{dep}[\bullet]$ are shown in Eq. (\ref{superoperators}), and $\rho_{d}(t)$ is the density matrix of the QB system during the self-discharging process.

We assume that the initial state of battery in the self-discharging process is the final state of battery $\rho_{d}(0)$ in the charging process. The energy at time $t$ reads
\begin{equation}
\label{disCenergy}
E_{d}(t)={\rm{Tr}}[H_{q}\rho_{d}(t)],
\end{equation}
and the $l_{1}$ norm of coherence during the self-discharging process is
\begin{equation}
C_{d}(t)=\sum_{i\neq j}\mid\rho_{d}(t)\mid.
\end{equation}
\section{The charging and self-discharging properties} \label{section3}
In this section, we investigate the charging and self-discharging performance of the QB under three environmental effects. We mainly consider the resonance regime of the transition frequency between resonator and qutrit ($|0\rangle\leftrightarrow|1\rangle$) unless mentioned otherwise, i.e., $\omega_{r}=\omega_{1}-\omega_{0}=\omega$ in Eq. (\ref{RQHam}). For simplicity, we treat all parameters are in units of $\omega$ and set $\omega=1, \omega_{0}=0, \omega_{1}=1, \omega_{2}=1.95, n_{rk}=2N, \Gamma_{12}=2\Gamma_{01}, and \Gamma_{22}=2\Gamma_{11}$.

\subsection{The stable charging with resonator decay and qutrit decoherence}
In the following, we focus on the charging performance of three QBs defined by Eq. (\ref{RQHam}), including one-mode, two-modes single-cell QB and one-mode many-body QB. The two single-cell QBs are defined as resonator-qutrit QBs consisting of a qutrit and a one-mode or two-modes resonator. In order to evaluate whether the single-cell QBs have charging advantages over existing QBs based on superconducting circuits, we also discuss the qutrit QB \cite{PhysRevE.100.032107}, which consists of a qutrit driven by two classical fields. The Hamiltonian of the qutrit QB is defined by
\begin{eqnarray}
H_{qutrit}=H_{0}+H_{t},
\end{eqnarray}
where
\begin{eqnarray}
\begin{split}
&H_{0}=\omega_{0}|0\rangle\langle0|+\omega_{1}|1\rangle\langle1|+\omega_{2}|2\rangle\langle2|,\\
&H_{t}=\Omega_{01}e^{-I\omega_{01}t}|0\rangle\langle1|+\Omega_{12}e^{-I\omega_{12}t}|1\rangle\langle2|+\rm{H.c.}.
\end{split}
\end{eqnarray}
Here $\omega_{mn}=\omega_{n}-\omega_{m}~(m, n=0,1,2)$ is the transition frequency from states $|m\rangle$ to $|n\rangle$. $\Omega_{01}(t)=\Omega_{0}f(t), \Omega_{12}(t)=\Omega_{0}[1-f(t)]$ are two classical fields, $\Omega_{0}$ is a parameter related to the amplitude of two fields, and $f(t)=t/\tau$ is a function which satisfies $f(0)=0$ and $f(\tau)=1$.
\begin{figure}[htbp]
\centering
\includegraphics[width=0.5\textwidth]{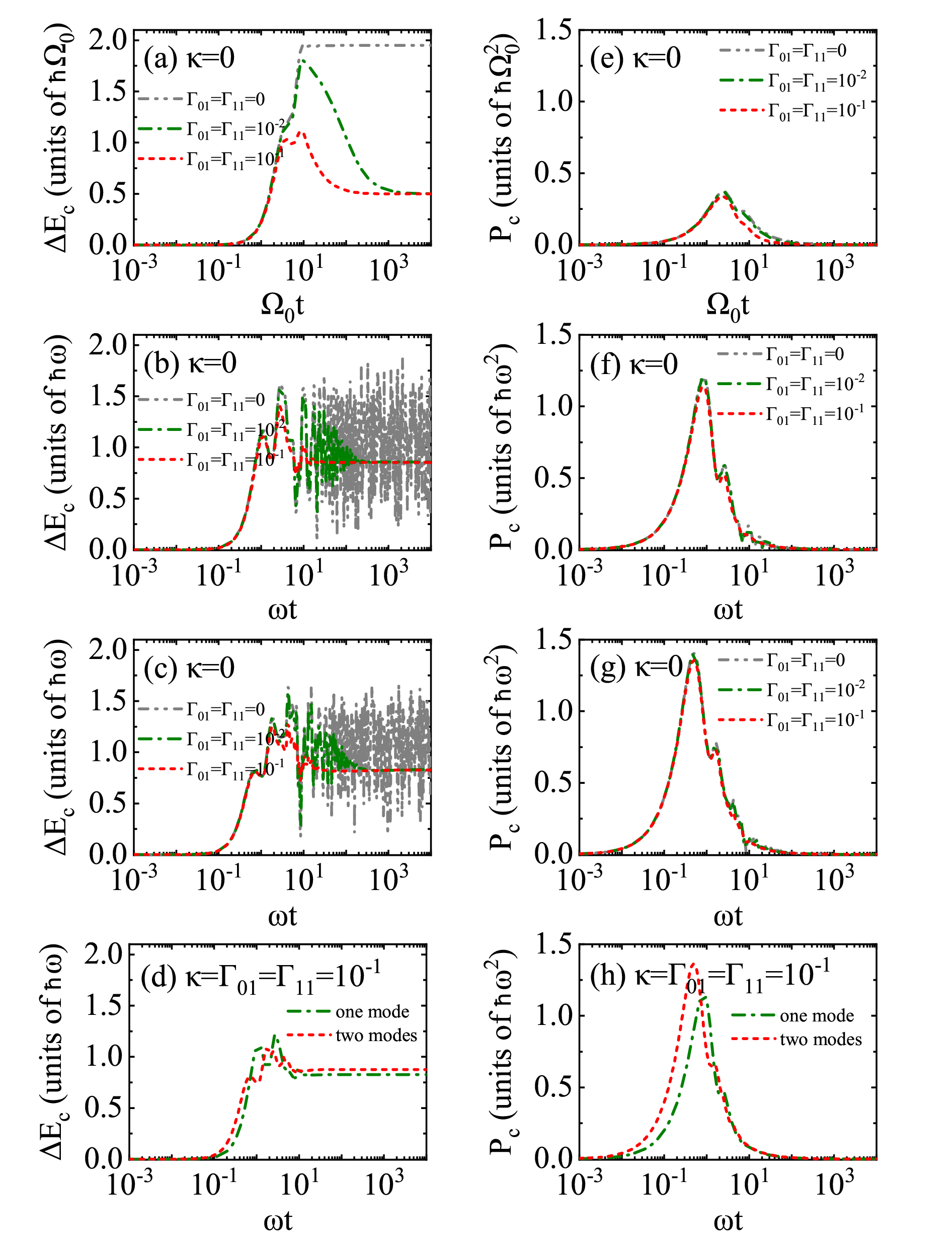}
\caption{(a)-(d) The time evolution of the stored energy $\Delta E_{c}(t)$ (units of $\hbar\Omega_{0}, \hbar\omega$) and (e)-(h) the average charging power $P_{c}(t)$ (units of $\hbar\Omega_{0}^{2},\hbar\omega^{2}$) in single-cell QBs with qutrit relaxation and dephasing, where (a) and (e) describe the qutrit QB, (b)-(d) and (f)-(h) describe the one-mode and two-modes resonator-qutrit QBs without and with resonator decay. The gray dash-dotted curve corresponds to the closed-system case, whereas other curves corresponds to the open-system cases. The parameters are chosen as $N=1, J=0$, and $g=1$.}
\label{fig1}
\end{figure}

The stored energy $\Delta E_{c}(t)$ of the qutrit QB and the resonator-qutrit QBs over the evolution time are shown in Fig. \ref{fig1} (a)-(d).
In closed systems, the stored energy of the resonator-qutrit QBs are unstable and highly oscillatory. However, the qutrit QB using stimulated Raman adiabatic passage inhibits the non-adiabatic excitation during the charging process, thus achieving a QB with stable and higher energy storage.
In open systems, the quantum interference suppresses the highly oscillatory phenomenon and results in the resonator-qutrit QBs eventually reaching steady states (subradiant states) \cite{Albrecht2019,PhysRevLett.124.013601,Zanner2022}. Comparing the red and green curves in Fig. \ref{fig1} (a)-(c), we find that the stable energy of the resonator-qutrit QBs are higher and this advantage still exists when considering the resonator decay of the resonator-qutrit QBs, see cases $\kappa\neq0$ in Fig. \ref{fig1} (d).
The time evolutions of the average charging power $P_{c}(t)$ in these single-cell QBs are described by Fig. \ref{fig1} (e)-(h). It is worth noting that the average charging power of the resonator-qutrit QBs are always significantly higher than that of the qutrit QB, regardless of whether open systems are taken into account. In particular, the average charging power of the resonator-qutrit QB is the highest when the two-modes resonator is used as a charger. This indicates that the resonator-qutrit QBs have more efficient charging processes than the qutrit QB, and the multiplied increasing photons in the two-modes resonator-qutrit QB leads to an acceleration mechanism of its charging process.

\begin{figure}
\centering
\includegraphics[width=0.5\textwidth]{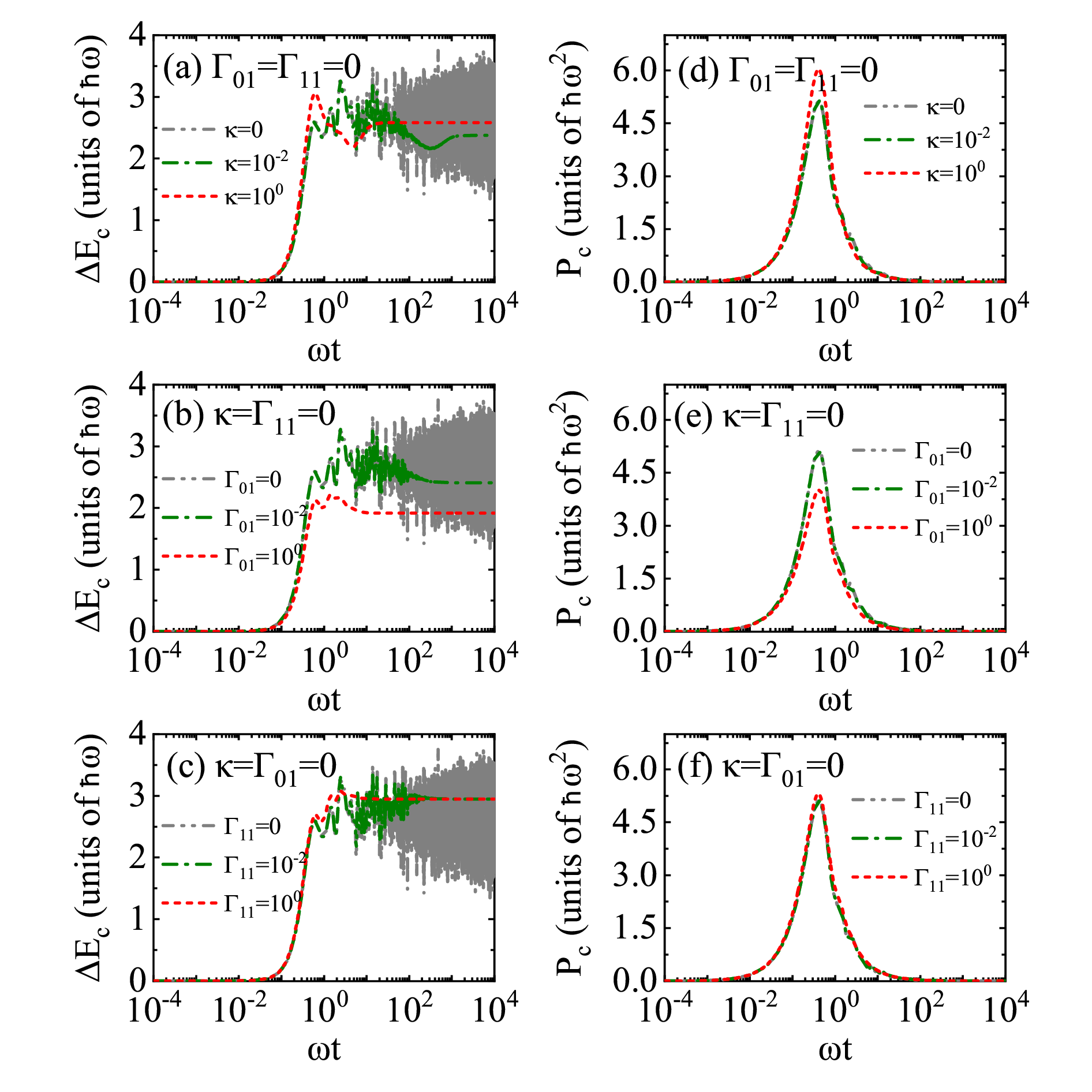}
\caption{(a)-(c) The time evolution of the stored energy $\Delta E_{c}(t)$ (units of $\hbar\omega$) and (d)-(f) the average charging power $P_{c}(t)$ (units of $\hbar\omega^{2}$) in many-body QB. The parameters are chosen as $N=3$, and $J=g=1$.}
\label{fig2}
\end{figure}
\begin{figure}
\centering
\includegraphics[width=0.46\textwidth]{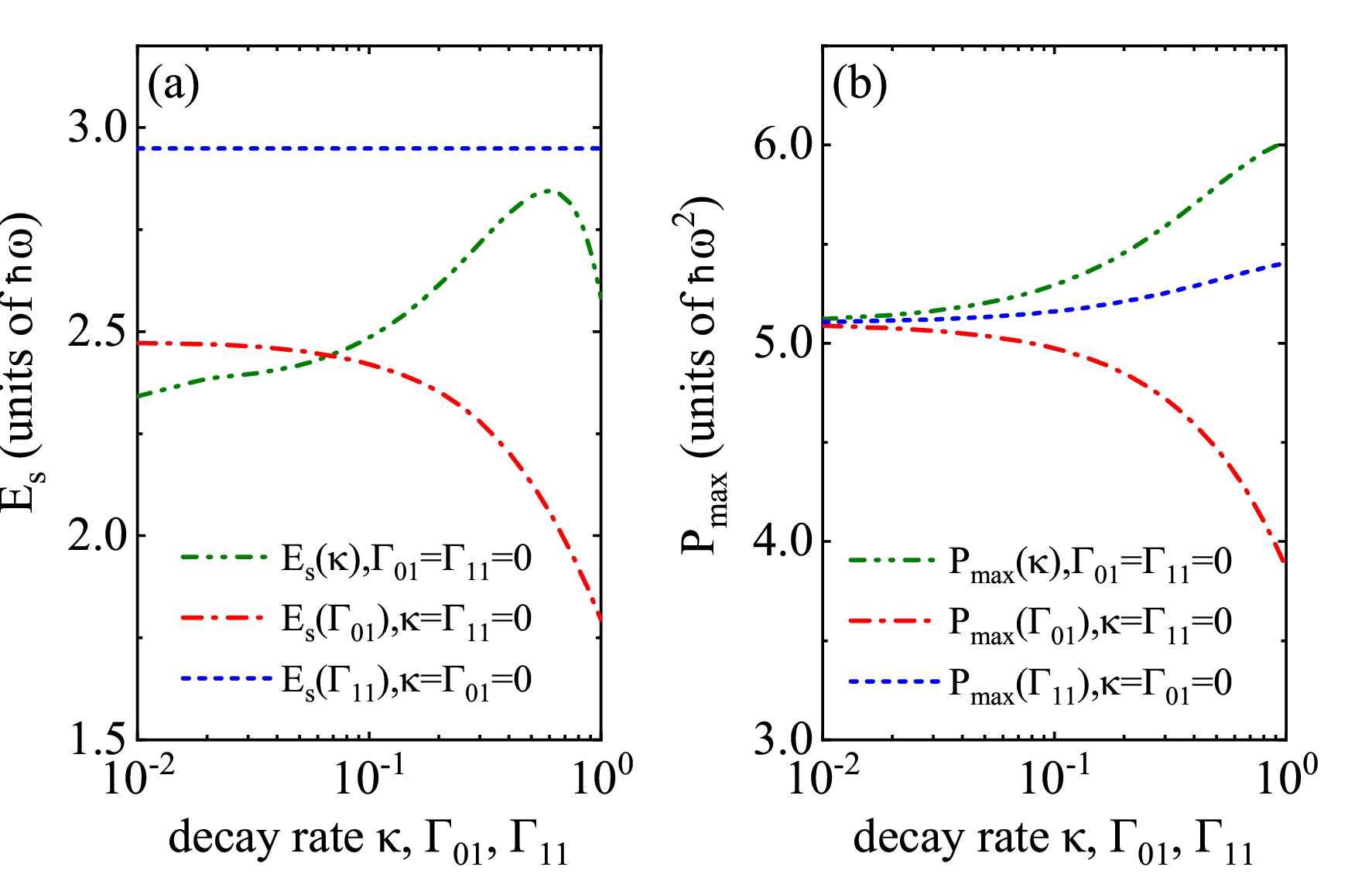}
\caption{The stable energy $E_s$ (in unit of $\hbar\omega$) and the maximum charging power $P_{max}$ (in unit of $\hbar\omega^{2}$) as a function of the decay rate (the resonator decay rate $\kappa$, the qutrit relaxation rate $\Gamma_{01}$ and the qutrit dephasing rate $\Gamma_{11}$). The other parameters are the same as in Fig. \ref{fig2}.}
\label{fig3}
\end{figure}
Although the stable energy of the resonator-qutrit QB is higher compared to the qutrit QB, its stored energy are still very limited. Next we will discuss the charging performance of the resonator-qutrits QB, namely many-body QB, which consists of a one-mode resonator and $N$ coupled qutrits.
As shown in Fig. \ref{fig2}, the stored energy $\Delta E_{c}(t)$ and the average charging power $P_{c}(t)$ of the resonator-qutrits QB can be obtained by solving Eq. (\ref{Ceq}). By utilizing subradiant states to tune qutrits into a decoherence-free subspace \cite{PhysRevLett.81.2594,Kwiat2000,PhysRevLett.120.140404}, the resonator-qutrits QB also have a stable charging process when considering the resonator decay and the qutrit decoherence.
In addition, Figure \ref{fig2} also indicates that the resonator decay has a positive effect on the stable and efficient charging of the resonator-qutrits QB, while the qutrit relaxation suppresses the population inversion between quantum states \cite{PhysRevA.76.042319,Krantz2019}, which leads to a remarkable decay on the stable energy and the maximum charging power of the resonator-qutrits QB.
The stable energy and the maximum charging power with decay rate in Fig. \ref{fig3} verify this conclusion. It is also observed that compared to the resonator decay and the qutrit relaxation, the stable energy of the resonator-qutrits QB is higher and with good robustness when considering the qutrit dephasing. The qutrit dephasing causes transitions between superradiant and subradiant states so that suppresses the emission of photons into environment. Meanwhile, compared with our previous work \cite{PhysRevA.107.023725}, we find that the resonator-qutrits QB can store higher energy and charge faster than the resonator-qubits QB, although two QBs both have efficient charging and stable energy-storage processes when considering the cases of open systems. The result demonstrates that the energy leakage to the second excited state can boost the performance of our QB.
\subsection{The slow self-discharging with qutrit decoherence}
Similar to classical batteries, QBs also have a phenomenon known as self-discharging process, which is characterized by the loss of energy in the battery even when no consumption hubs are coupled to them \cite{Kamin2020,PhysRevE.100.032107,PhysRevA.104.042209,PhysRevE.103.042118,PhysRevE.105.054115}. The self-discharging process is associated with undesired effects that deteriorate the energy-storage performance of the battery \cite{Kamin2020}. Next, we will mainly explore how to reduce the speed of self-discharging process of the resonator-qutrits QB.

We first consider the case of $N=1$ in Eq. (\ref{disCeq}) and express its Hamiltonian and density matrix during the self-discharging process as
\begin{equation*}
H_{q}=\begin{bmatrix}
      \omega_{0}&0&0\\
      0&\omega_{1}&0\\
      0&0&\omega_{2}
      \end{bmatrix},
\rho_{d}=\begin{bmatrix}
         \rho_{11}&\rho_{12}&\rho_{13}\\
         \rho_{21}&\rho_{22}&\rho_{23}\\
         \rho_{31}&\rho_{32}&\rho_{33}
         \end{bmatrix}.
\end{equation*}
The energy defined in Eq. (\ref{disCenergy}) can be rewritten as
\begin{equation}
E_{d}(t)=\omega_{0}\rho_{11}(t)+\omega_{1}\rho_{22}(t)+\omega_{2}\rho_{33}(t).
\end{equation}
Since $E_d(t)$ depends only on the diagonal element of density matrix $\rho_{d}$, the problem of finding the analytic solution of $E_d(t)$ reduces to the task of solving the following equations
\begin{equation}
\begin{split}
&\dot{\rho_{11}}(t)=\Gamma_{01}\rho_{22}(t),\\
&\dot{\rho_{22}}(t)=-\Gamma_{01}\rho_{22}(t)+\Gamma_{12}\rho_{33}(t),\\
&\dot{\rho_{33}}(t)=-\Gamma_{12}\rho_{33}(t),
\end{split}
\end{equation}
whose solution is as follows
\begin{equation}
\begin{split}
&\rho_{11}(t)=\frac{\Gamma_{12}e^{-\Gamma_{01}t}-\Gamma_{01}e^{-\Gamma_{12}t}}{\Gamma_{01}-\Gamma_{12}}+1,\\
&\rho_{22}(t)=\frac{\Gamma_{12}e^{-\Gamma_{12}t}-\Gamma_{12}e^{-\Gamma_{01}t}}{\Gamma_{01}-\Gamma_{12}},\\
&\rho_{33}(t)=e^{-\Gamma_{12}t},
\end{split}
\end{equation}
where we assume the initial state of the battery during the self-discharging process in its second excited state $|2\rangle$ and use the normalization condition $\rho_{11}(t)^{2}+\rho_{22}(t)^{2}+\rho_{33}(t)^{2}=1$. The analytic solution of the energy during the self-discharging process is as follows
\begin{equation}
\label{disCenergysolution}
E_{d}(t)=\frac{e^{-\Gamma_{12}t}(\Gamma_{01}\omega_{02}-\Gamma_{12}\omega_{12})-e^{-\Gamma_{01}t}\Gamma_{12}\omega_{01}}{\Gamma_{01}-\Gamma_{12}}+\omega_{0}.
\end{equation}

\begin{figure}[htbp]
\centering
\includegraphics[width=0.46\textwidth]{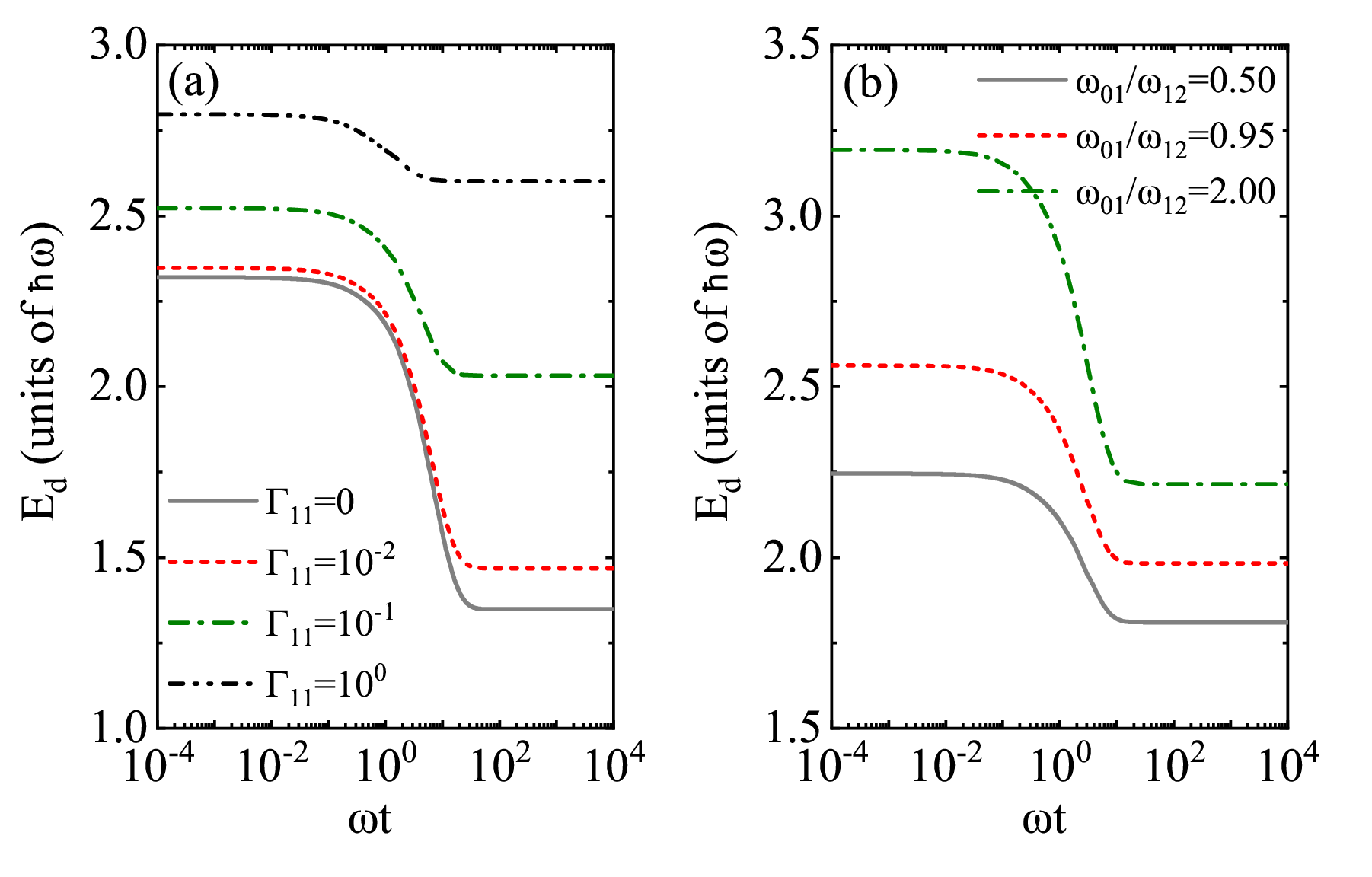}
\caption{The time evolution of the energy $E_{d}(t)$ (units of $\hbar\omega$) in the self-discharging process of the resonator-qutrits QB for different choices of the dephasing rates $\Gamma_{11}$ and the energy gaps $\omega_{01}/\omega_{12}$. The parameters are fixed as $N=3, g=J=1$, and $\kappa=\Gamma_{01}=10^{-1}$.}
\label{fig4}
\end{figure}

For the single-cell QB, Eq. (\ref{disCenergysolution}) reveals that the energy during the self-discharging process is only affected by the energy gap and relaxation rate, not related to the dephasing rate. However, for the resonator-qutrits QB, due to the presence of nearest neighbor interaction terms between qutrits, the dephasing process inevitably affects its self-discharging process. Therefore, we fix the relaxation rate and focus on the effects of the dephasing rate and the energy gap on the energy during the self-discharging process of the resonator-qutrits QB, as shown in Fig. \ref{fig4}. It is obvious that the energy stored in the resonator-qutrits QB increases and the self-discharging speed decreases with the increase of the dephasing rate and the relative energy gap. This means that we can realize a longer-lived QB with both an efficient charging process and a slow self-discharging process by considering the dephasing process and manipulating the energy gap.
\section{The roles of quantum entanglement and quantum coherence} \label{section4}
As the QB charges superextensively, it would also discharge superextensively if it were simply disconnected from the charger \cite{PhysRevApplied.14.024092}. However, our  
  resonator-qutrits QB has both an efficient charging process and a slow self-discharging process. This is because we propose a dephasing process before disconnecting the charger. In this section, in order to find out how the dephasing process specifically determines the behavior of the energy in the resonator-qutrits QB, we introduce coherence and entanglement, characterized by the $l_{1}$ norm of the off-diagonal elements and the logarithmic negativity, respectively.

\begin{figure}[htbp]
\centering
\includegraphics[width=0.5\textwidth]{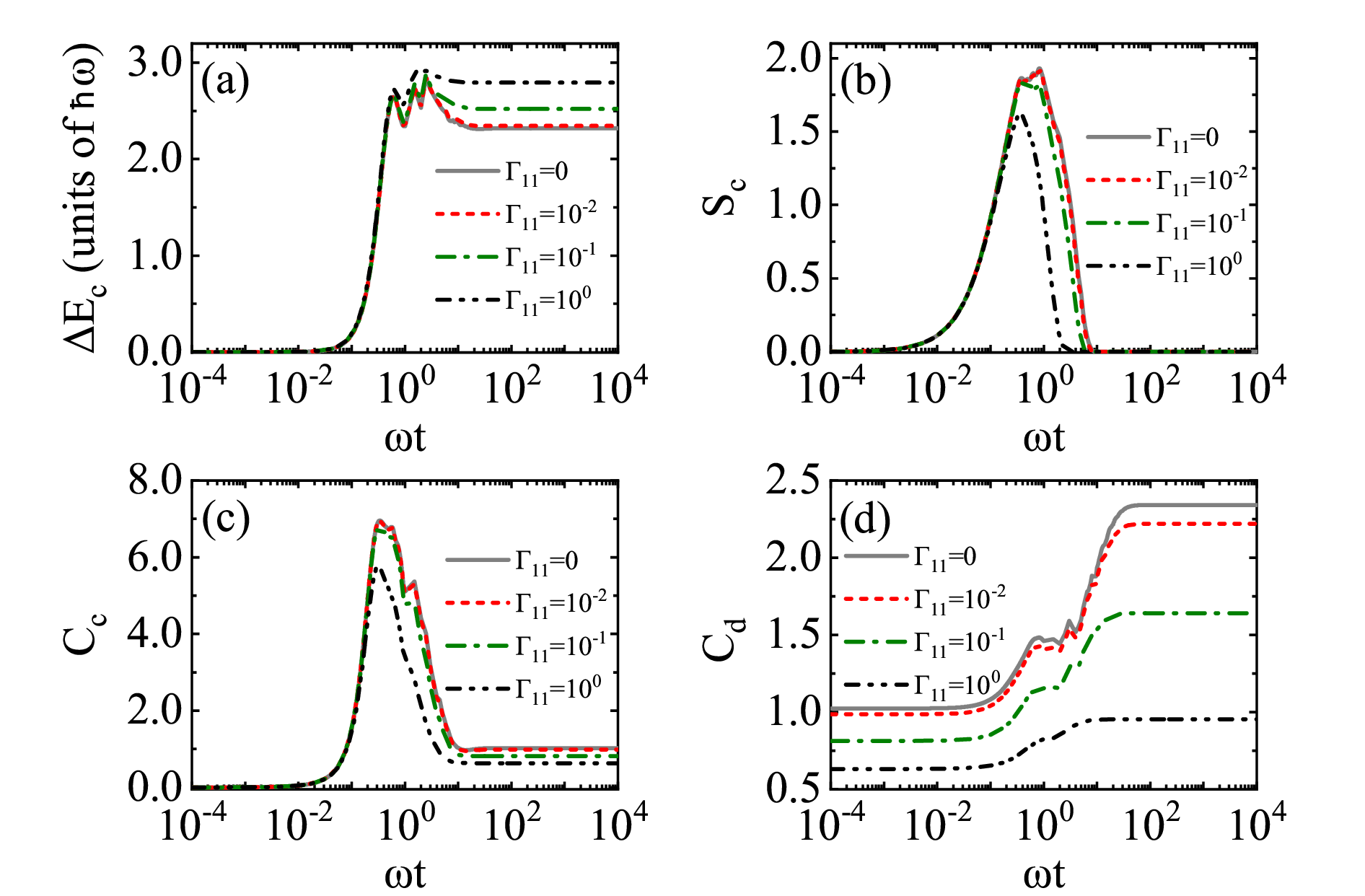}
\caption{(a)-(b) Entanglement and energy: The time evolution of the stored energy $\Delta E_{c}(t)$ (units of $\hbar\omega$) and the logarithmic negativity $S_{c}$ during the charging process of the resonator-qutrits QB. (c)-(d) Coherence: The time evolution of the $l_{1}$ coherence $C_{c}(t), C_{d}(t)$ during the charging and self-discharging processes of the resonator-qutrits QB. The other parameters are the same as in Fig. \ref{fig4}.}
\label{fig5}
\end{figure}
In Fig. \ref{fig5}, we illustrate the evolutions of entanglement and coherence for different dephasing rates. Before the maximum energy is reached during the charging process, note that if there is no battery's coherence and no battery-charger entanglement, then the energy of the resonator-qutrits QB is always zero. This means the energy is positively related to coherence and entanglement and further demonstrates that the coherence in the battery or the entanglement between battery and charger is a necessary resource for generating nonzero energy during the charging process.
At the end of the charging process, we can establish a tight link of coherence and entanglement with the stable energy: (\romannumeral1) The relationship between coherence and energy shows that lower coherence corresponds to higher steady-state energy, meaning that the coherence in battery inhibits the stable energy of the resonator-qutrits QB. (\romannumeral2) The phenomenon of the battery-charger entanglement suddenly disappearing when the battery reaches steady states just verifies the physical mechanism of steady-states generation, that is, the dynamic decoupling behaviour due to the quantum interference caused by the collective effects between the QB and the environment.
In addition, during the whole self-discharging process, it is worth mentioning that the increasing coherence causes the battery to discharge at a superradiantly decay rate and has a detrimental effect on the energy, as shown in Fig. \ref{fig4} (a) and Fig. \ref{fig5} (d). Thus, our results suggest that the dephasing process, which destroys the coherence of the battery in its energy eigenbasis, shows a counterintuitive advantage in the self-discharging process of the resonator-qutrits QB.
\section{Conclusions} \label{section5}
In this work, we have derived the Hamiltonian of a light-matter interaction model based on superconducting circuits platform and defined three QBs.
We have investigated the charging performance of these QBs in the presence of the resonator decay and the qutrit decoherence. Our results showed that, for single-cell case, the one-mode or two-modes resonator-qutrit QBs have both higher stable energy and charging power than existing QBs based on superconducting circuits. For many-body case, we have achieved a resonator-qutrits QB with a stable and efficient charging process by utilizing subradiant states to tune qutrits into a decoherence-free subspace.
Meanwhile, we have explored how to reduce the speed of self-discharging process of the resonator-qutrits QB. Our findings suggested that the resonator-qutrits QB with a slow self-discharging process can be realized by considering the dephasing process and manipulating the energy gap.
Remarkably, we have also emphasized the necessity of battery's coherence and battery-charger entanglement for generating nonzero energy of the resonator-qutrits QB. The charing energy is positively related to coherence and entanglement while the stable energy and the self-discharing energy are negatively related to coherence. The phenomenon of entanglement suddenly disappearing when the battery reaches steady states corresponds to the dynamic decoupling behavior caused by quantum interference. Our QBs address inadequacies of the previously proposed QBs based on superconducting circuits systems \cite{Hu2022,Zheng2022,Ge2023}. The results provide an alternative way for further realization of many-body QBs with efficient charging, stable energy-storage and slow self-discharging processes on superconducting circuits platform.
\section*{Acknowledgments}
The work is supported by the National Natural Science Foundation of China (Grant No. 12075193).
\bibliography{refercence}

\end{document}